\documentclass{article}

\usepackage{graphicx}
\usepackage{graphics}
\usepackage{amsmath,amssymb,amsfonts}
\usepackage{colortbl}
\usepackage{array}
\usepackage{bm}
\usepackage[style=numeric,sorting=none]{biblatex}
\newcommand*{\Hc}{\textrm{H.c.}}
\newcommand*{\ra}{\rangle}
\newcommand*{\la}{\langle}
\newcommand*{\up}{\uparrow}
\newcommand*{\dn}{\downarrow}
\newcommand{\A}{\mathbf{A}}
\newcommand*{\eps}{\varepsilon}
\newcommand*{\R}{\mathbf{R}}

\renewcommand{\a}{\mathbf{a}}

\newcommand*\bsigma{\bm{\sigma}}

\begin{document}
\title{Light-matter interaction across a Weyl-Kondo semimetal to Kondo insulator topological phase transition}
\maketitle

 Jason Spada$^{1}$, Anthony Acebal$^{1}$, Emiliano Benitez$^{1}$,  Benedikt Fauseweh$^2$ and Jian-Xin Zhu$^{3}$, Sarah E. Grefe$^{1,*}$

$^1$\textit{Department of Physics and Astronomy, California State University Long Beach, Long Beach, CA, USA}

$^2$\textit{Department of Physics, TU Dortmund University, Dortmund, Germany}

$^3$\textit{Theoretical Division 4, Los Alamos National Laboratory, Los Alamos, NM, USA}

$^*$\textit{Author to whom any correspondence should be addressed: sarah.grefe@csulb.edu}


\begin{abstract}

For strongly correlated materials such as the Weyl-Kondo semimetal and Kondo insulator, conventional tools to measure band geometry can fail to give a clear picture of linear dispersion or small band gaps, respectively. 
There is a need for new probes that work at small energy resolution in the THz regime of quantum materials.
In this work, we calculate the high-harmonic spectroscopy, direct-current response, and photo-induced post-pulse states of the topological phase transition between Weyl-Kondo semimetal and Kondo insulator states.
We develop a non-perturbative theoretical framework to treat the current response to a driving field that includes all orders of the field.
We find that there are substantial qualitative and quantitative signatures distinguishing the two states, in particular that although no symmetry is broken through the topological phase transition, the gapped state can be detected in the high-harmonic spectroscopy and direct-current response.
We also find that pulsed light is able to influence the heavy fermion order parameters in the Weyl-Kondo semimetal to a distinct post-pulse state.

\end{abstract}

\section{Introduction}

The interplay of topology and strong electron correlations is a rising topic in condensed matter physics.
On one hand, topological states are associated with topological invariants that unmistakably identify states as either topologically trivial or topologically nontrivial, yet depending on the symmetry of the system may or may not be an observable quantity.
On the other hand, strong correlations generally lead to a narrowing of the energy bands near the Fermi surface due to renormalization and many-body resonances, as well as a smearing out of the bands with respect to their resolution in momentum space.
This makes the observation of strongly correlated topological states challenging with traditional tools such as angle-resolved photoemission spectroscopy (ARPES), such as the strongly renormalized linearly dispersing bands of Weyl-Kondo semimetals (WKSMs)~\cite{Lai2018,Grefe2020} in candidate materials such as Ce$_{3}$Bi$_{4}$Pd$_{3}$~\cite{Dzsaber2017,Dzsaber2021,Dzsaber2022}.
In Ce$_{3}$Bi$_{4}$Pd$_{3}$, a linear dispersion attributed to Weyl fermions develops at temperatures below the Kondo temperature $T_K$, corresponding to renormalized bands in an energy window of $k_BT_K\approx1$meV, in the THz range~\cite{Dzsaber2017}.
Probing such a material would require tools with sub-meV energy resolution.
This exposes the need for alternative probes of topology, band structure, and order parameters in strongly correlated topological phases.

Solid-state high-harmonic generation (HHG) was established in bulk ZnO and is understood as an intraband current plus an interband polarization, with the spectrum encoding crystal orientation~\cite{Ghimire2011,Vampa2014,Schubert2014,You2017} and inversion-breaking even harmonics~\cite{Liu2017,Ghimire2019}. 
For circularly polarized drives, a crystal with an \(N\)-fold rotation axis obeys a dynamical symmetry that restricts emission to orders \(qN \pm 1\) with alternating handedness~\cite{Alon1998,Neufeld2019,Saito2017}.
In Weyl and Dirac semimetals, terahertz harmonics were measured in Cd\(_3\)As\(_2\) and infrared-driven harmonics in \(\beta\)-WP\(_2\), while a first-principles TaAs study concluded that harmonic observables follow the symmetries shared by the pulse and crystal rather than the Weyl nodes~\cite{Kovalev2020,Lv2021,Zhang2024}. 
Related work studied surface-state harmonics in three-dimensional topological insulators~\cite{Schmid2021}, edge states in a one-dimensional model~\cite{Bauer2018}, helicity and circular dichroism in the Haldane model~\cite{Silva2019,Chacon2020}, and the response to a circular drive across a topological phase transition in Bi\(_2\)Se\(_3\)~\cite{Heide2022}.
Topological signals include circular dichroism tracking the Chern number and helicity switching in the Haldane model~\cite{Chacon2020,Silva2019}, though in Weyl semimetals the TaAs study indicates polarization rules are governed by crystal symmetry rather than the nodes themselves~\cite{Zhang2024}.

HHG in strongly correlated systems has mostly focused on Mott insulators, where doublon-holon dynamics yield a cutoff that can be set by the interaction in the strong-field regime rather than by the band gap, with an experiment on Ca\(_2\)RuO\(_4\)~\cite{Silva2018,Murakami2018,Uchida2022}. 
Correlation signatures include doublon-holon dynamics and an interaction-set harmonic cutoff~\cite{Silva2018,Murakami2018}.
For Kondo systems, a prior theoretical study of a laser-driven one-dimensional Kondo lattice includes high-harmonic spectroscopy, and found that the Kondo hybridization gap could be collapsed with a critical intensity of light~\cite{Zhu2021}.

DC field-driven work has been studied on a variety of quantum materials such as Mott insulators, topological insulators, quantum spin liquids, and WKSMs. 
DC fields have been used to control topological orders, manipulate topological phase transitions, and observe topological responses.
Recent work on Mott insulators has proposed the control of magnetic and topological orders using a DC electric field~\cite{Takasan2019}. 
Electric field-driven topological phases have also been observed in Kitaev quantum spin liquids, where increasing the amplitude of the electric field induced a topological phase transition between critical states and bulk energy-gapped states~\cite{Noh2024}. 
Furthermore, the hybridized BiSbTeSe$_{2}$ three-dimensional topological insulator has shown a phase transition between a trivial and nontrivial topological phase, in which a transverse electric field is used to realize a quantum spin insulator with a gapless surface state and an emerging edge conductance state of a normal insulator~\cite{Chong2022}.
The WKSM candidate material Ce$_{3}$Bi$_{4}$Pd$_{3}$ shows a spontaneous Hall effect driven by a DC current observed below 3 K with zero applied magnetic field~\cite{Dzsaber2017,Dzsaber2021}.

Photo-induced phase transitions have also been studied as a way to detect and change the properties of materials~\cite{Bennemann2011}. 
Previous work on quantum materials used an ultrafast IR pump method to photo-induce a structural phase transition in the strongly correlated VO$_2$~\cite{Tao2016}. 
Weyl semimetals have also been shown to have photo-induced topological phase transitions, from a line-crossing nodal semimetal to a multiple Weyl semimetal~\cite{Ezawa2017}. 
Photo-induced quantum phase transitions have been observed in strongly correlated systems such as the Hubbard Model, where the time evolution of the spin-orbit order parameters were tracked as a pulse is applied, resulting in the system settling into a hidden non-equilibrium state~\cite{Li2018}.

In this work, we study the WKSM to Kondo insulator (KI) topological phase transition using three methods of study: HHG, DC response, and photoinduced phase transitions.
The purpose of this work is to compare topologically trivial and topologically nontrivial phases in a strongly correlated model using a variety of tools that can potentially be tested in materials platforms such as Ce$_{3}$Bi$_{4}$Pd$_{3}$ (a WKSM candidate) and Ce$_{3}$Bi$_{4}$Pt$_{3}$ (a well-known KI), that go beyond the energy resolution limitations of conventional tools such as ARPES.
We find that the various responses are qualitatively and quantitatively different between WKSM and KI phases due to band structure differences and degree of spin orbit coupling.
We also find that an $\omega_0=0$ pulse is able to influence the heavy fermion order parameters in the WKSM system.

First, we simulate few-cycle circularly and linearly polarized pulses driving the WKSM model to determine which features of the harmonic spectrum track the staggered-mass-driven annihilation of Weyl nodes pinned near the Fermi energy into the KI state~\cite{Lai2018,Grefe2020,Dzsaber2017}.
There are quantitative and qualitative differences across the topological phase transition.

Second, we study the DC current response across the phase transition using zero frequency field or DC field excitation. 
We find that the DC response is absent in the KI state (there is no photo-induced metallicity) whereas there is a finite response in the WKSM state, which holds true for a range of  DC field strengths and direction of field orientations.
We also find that the WKSM and KI show marked differences in their DC response phase diagram of spin-orbit coupling strength versus intensity.

Finally, in this work we simulate an ultra fast laser pulse to affect the Kondo hybridization order parameter (the auxiliary boson amplitude $r$) and track its dynamics.
We find that the strength of the driving field is proportional to the suppression of the auxiliary boson to a post-pulse state.
This hints that for future work, total suppression of the auxiliary boson may drive the system to Kondo destruction and photo-induce a quantum phase transition~\cite{Si2013}. 

The paper is organized as follows. 
In Section~\ref{sec:methods}, we introduce the equilibrium Hamiltonian (Sec.~\ref{sec:eqH}) of the model which yields the WKSM and KI phases, then modify it with a vector potential field through the Peierls transformation (Sec.~\ref{sec:peierls}) to accommodate a generic excitation field, and then detail how the states are time-evolved in Sec.~\ref{sec:time}.
We introduce a generalized multi-orbital current operator in Sec.~\ref{sec:current} which includes all harmonic orders of the excitation field of frequency $\omega_0$.
In the Results section (Sec.~\ref{sec:results}), we show simulated HHG spectra for linearly polarized light (Sec.~\ref{sec:omega_lin}) and circularly polarized light (Sec.~\ref{sec:omega_circ}).
We also show phase diagrams of the DC excitation and response ($\omega_0=0$) of the WKSM versus KI in Sec.~\ref{sec:omega_DC}, then present the dynamics of the heavy fermion order parameters when excited by a short $\omega_0=0$ pulse in Sec.~\ref{sec:rt_lt}.
Lastly, we discuss the implications of our work in Section~\ref{sec:dis}.

\section{Methods}\label{sec:methods}

To show how we simulate the HHG responses of the WKSM and KI phases, we first introduce the equilibrium Hamiltonian in section~\ref{sec:eqH}, then its Peierls substitution in ~\ref{sec:peierls} which allows the kinetic terms in the Hamiltonian to be affected by the applied vector potential field.
In Section~\ref{sec:time} we detail the Schr\"{o}dinger equation-based time evolution method, and in Section~\ref{sec:current} we explain our non-perturbative expression for the current which is generic enough to contain all orders of the excitation electric field.

\subsection{Equilibrium Hamiltonian}\label{sec:eqH}
The equilibrium Hamiltonian is given by
\begin{align}
	\mathcal{H} &= \mathcal{H}_c + \mathcal{H}_{cd} + \mathcal{H}_{d}\label{eq:anderson}\\
	\mathcal{H}_c 
	&= t\sum_{\la i j \ra, \sigma } \left( c_{i \sigma}^\dagger c_{j \sigma}+ \Hc \right)
	+\sum_{i, \sigma}\left(m(-1)^i - \mu\right) n_{i \sigma}^c + i \lambda \sum_{\la\la i j \ra\ra} \left[ c^\dagger_{i \sigma} \left( \bsigma\cdot {\bf e}_{ij} \right)_{\sigma\sigma'} c_{j\sigma'} - \Hc\right]\label{eq:hc}\\
	\mathcal{H}_{cd} &=
V \sum_{i,\sigma} \left( d_{i \sigma}^\dagger c_{i \sigma} + \Hc\right)\label{eq:hcd}\\
	\mathcal{H}_d&=\sum_{i, \sigma}(E_d-\mu) d_{i \sigma}^\dagger d_{i \sigma}+ U \sum_i n^d_{i \up} n^d_{i \dn}\label{eq:hd}.
\end{align}
The WKSM model is an Anderson lattice Hamiltonian situated on the diamond lattice structure, consisting of three main parts as in equation \ref{eq:anderson}: a part affecting just the conduction electrons $\mathcal{H}_c$, a part affecting just the localized $f$-orbital electrons in $\mathcal{H}_d$, and a hybridization term between the two species in $\mathcal{H}_{cd}$.
The conduction electron term $\mathcal{H}_c$ contains nearest-neighbor hopping with amplitude $t$, a sublattice-differentiating inversion symmetry breaking term with strength $m$, a chemical potential $\mu$, and a Dresselhaus-type spin orbit coupling parametrized with $\lambda$. 
In the spin-orbit coupling term, ${\bf e}_{ij}={\bf e}_{i}\times{\bf e}_{j}/|{\bf e}_{i}\times{\bf e}_{j}|$, where the ${\bf e}_{i}$ are lattice bond vectors connecting a site to its next-nearest neighbor.
The $f$-electron term contains a localized atomic-like energy level $E_d$, the chemical potential, and an on-site Coulomb interaction term with strength $U$.

We evaluate the strong coupling limit ($U\rightarrow\infty$) of this model to approximate the mixed-valence regime of heavy fermion systems, allowing us to use the auxiliary boson treatment~\cite{Coleman1984, Read1983, Newns1987, Millis1987, Coleman1987}.
In this treatment, we transform $d_{i \sigma}^\dagger \rightarrow f_{i \sigma}^\dagger b_i^\dagger$ and $d_{i \sigma} \rightarrow f_{i \sigma} b_i$, and use a Lagrange multiplier $\ell$ to fix the particle number of $f$-electrons.
We also approximate the bosons as a spatially invariant mean field $\langle b_i\rangle\rightarrow r$.
This transforms Eqs.~\ref{eq:hcd} and \ref{eq:hd} into:
\begin{align}
		\mathcal{H}_{cf} &=
rV \sum_{i,\sigma} \left( f_{i \sigma}^\dagger c_{i \sigma} + \Hc\right)\\
	\mathcal{H}_f&=\sum_{i, \sigma}(E_d-\mu) f_{i \sigma}^\dagger f_{i \sigma} + \ell \sum_{i,\sigma} (f_{i \sigma}^\dagger f_{i \sigma} + r^2 -1).
\end{align}

The auxiliary boson parameters $r$ and $\ell$, along with $\mu$ can be solved for self-consistently, given input values for the parameters $t,~m,~\lambda,~E_d,$ and $V$.
In equilibrium, the order parameters can be found by finding the roots of:
\begin{align}
	\left\langle\frac{\delta\mathcal{H}({\mathbf{k}})}{\delta\ell}\right\rangle &= r^2-1+\frac{1}{2N_u}\sum_{n,{\mathbf{k}}} \sum_{i=5}^{i=8} \langle\phi_{in}^*({\mathbf{k}}) f(\eps_{n{\mathbf{k}}})\phi_{in}({\mathbf{k}})\rangle=0,\label{eq:sc_r}\\
	\left\langle\frac{\delta\mathcal{H}({\mathbf{k}})}{\delta r}\right\rangle &= r\ell + \frac{V}{2N_u}\sum_{n,{\mathbf{k}}}\sum_{i=1}^{i=4} f(\eps_{n{\mathbf{k}}})\langle \phi_{in}^*({\mathbf{k}}) \phi_{(i+4)n}({\mathbf{k}})\rangle=0,\label{eq:sc_l}\\
	N_\text{tot}(\mu)&=\frac{1}{2N_u}\sum_{n,{\mathbf{k}}}\sum_{i=1}^{i=8} \langle\phi_{in}^*({\mathbf{k}}) f(\eps_{n{\mathbf{k}}})\phi_{in}({\mathbf{k}})\rangle -1=0,\label{eq:ntot}
\end{align}
as functions of $r,~\ell$, where $f(\eps_{n{\mathbf{k}}})$ is the Fermi-Dirac function at equilibrium.
The matrix elements indexed by $i$ are ordered such that $i=1,2,3,4~$ correspond to the conduction electron basis and $i=5,6,7,8~$ correspond to the $f$ electrons.
The total density $N_{\text{tot}}$ in Eq.~\ref{eq:ntot} is solved by the bisection method for the chemical potential $\mu$.
We solve the Eqs.~\ref{eq:sc_r},\ref{eq:sc_l} by Newton-Raphson method, which solves a coupled nonlinear set of equations.

\begin{figure}
 \centering
        \includegraphics[width=0.75\textwidth]{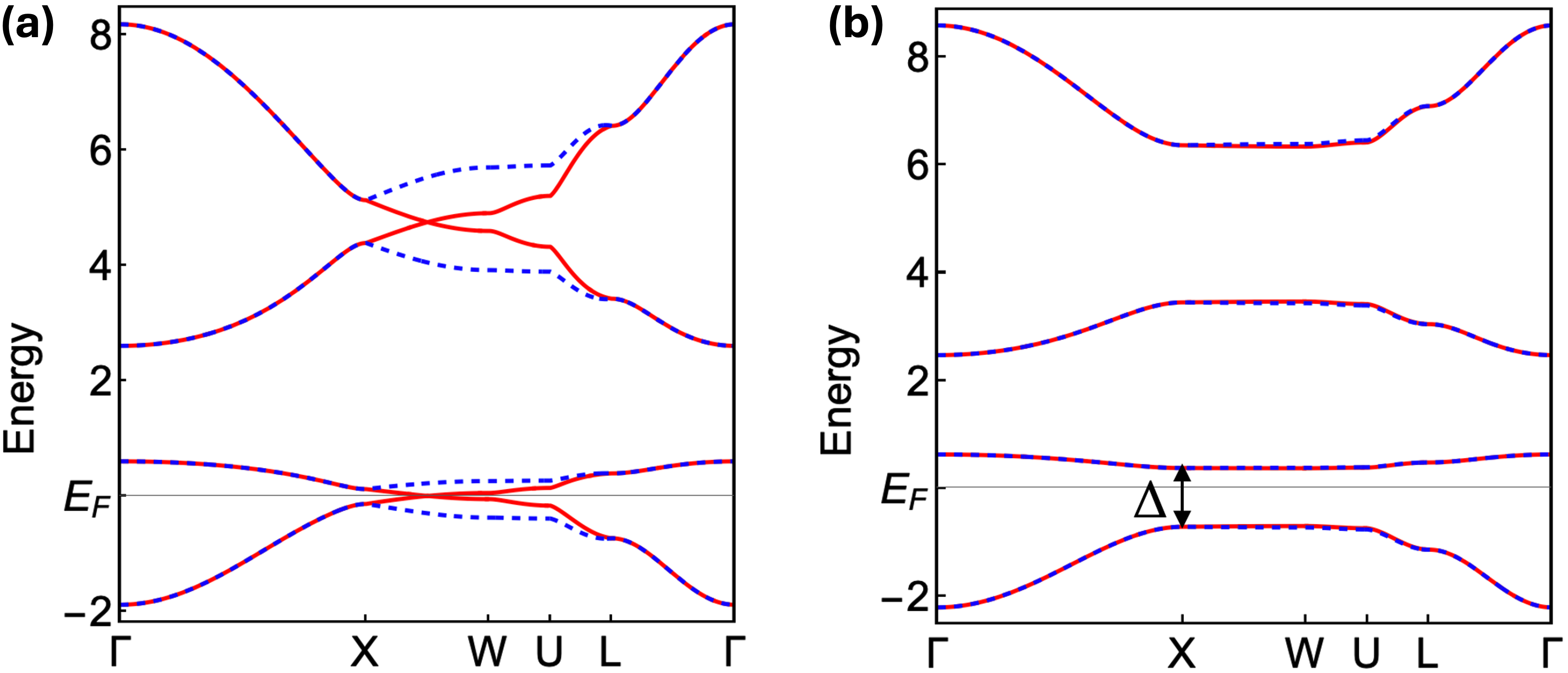}
 \caption{Equilibrium eigenenergy dispersions along a high-symmetry contour of the fcc Brillouin zone corresponding to the diamond lattice structure with on-site electrostatic potentials differentiating A and B sublattices (broken inversion symmetry). Both share the parameters $E_d=-4.5$ ($f$-electron level), $V=3$ (hybridization), $\mu\approx-3.53$ (chemical potential), $r^2=0.477$ (auxiliary boson field), $\ell=2.138$ (Lagrange multiplier).
(\textbf{a}) The Weyl-Kondo semimetal phase displaying linearly dispersing Weyl nodes pinned to the Fermi energy, with parameters $m=0.5$ (on-site sublattice potential difference), and $\lambda=0.25$ (spin-orbit coupling).
(\textbf{b}) The Kondo insulator phase displaying a gap of size $\Delta=1.088$ about the Fermi energy, with parameters $m=2$ (on-site sublattice potential difference), and $\lambda=0.01$ (spin-orbit coupling).
\label{fig:disp}}
\end{figure}

Importantly, a tuning between the topologically nontrivial WKSM phase and a topologically trivial Kondo insulator (KI) phase is achieved by tuning the ratio of $m/8\lambda$: When $0<m/8\lambda<1$, the phase is WKSM, and when $m/8\lambda>1$, the phase is KI.
This result can be seen in the eigenenergy dispersion in Fig.~\ref{fig:disp}, where the WKSM phase displays linearly-dispersing Weyl nodes between the $X$ and $W$ points of the Brillouin zone, resulting in 6 pairs of Weyl nodes, with pairs of opposite monopole chirality, throughout the Brillouin zone and related to each other by symmetry.
Tuning the parameters $m$ and $\lambda$ within $0<m/8\lambda<1$ moves the ${\mathbf{k}}$-space location of the Weyl nodes between $X$ and $W$.
The Weyl nodes are fixed to the Fermi energy $E_F$ as a result of Kondo-pinning, whereby the Kondo resonance combines with the space group symmetry constraints at quarter conduction electron filling per unit cell.
On the other hand, the KI phase displays a gap in the spectrum at the Fermi energy of size $\Delta$, which is mostly determined by the size of $m$.

\subsection{Peierls Hamiltonian}\label{sec:peierls}
In this section, we derive the Peierls-substituted Hamiltonian's kinetic terms, uncovering an important contribution to the nearest-neighbor hopping as a consequence of the diamond crystal structure's two sites per unit cell, and the form of the contribution to the spin-orbit coupling term. 
Given the vector potential for a light pulse,
\begin{align}
	\A(t)&=\left(A_x(t),~A_y(t),~A_z(t)\right),
\end{align}
it is assumed that the vector potential varies slowly on the scale of the unit cell length $a$, so its expression is spatially independent.
Each term of the kinetic energy operator representing a hopping between atomic sites is multiplied by a phase factor
\begin{align}
	c_{i\sigma}^\dagger&\rightarrow c_{i\sigma}^\dagger\exp\left(\tfrac{ie}{\hbar c}\int^{\R_i}_{\R_\infty} d\R\cdot\A(t)\right),\\
	c_{i\sigma}&\rightarrow c_{i\sigma}\exp\left(-\tfrac{ie}{\hbar c}\int^{\R_i}_{\R_\infty} d\R\cdot\A(t)\right)
\end{align}
where the integral path is a vector connecting two lattice site positions, and herein units of $\frac{e}{\hbar c}=1$ are used. 
	Applying the substitution to the nearest-neighbor hopping, and second-nearest-neighbor spin-orbit coupling of the Hamiltonian's conduction electrons, the phase factors become
\begin{align}
	t&\rightarrow t\exp\{-i\A(t)\cdot(\R_i-\R_j) \}\\
\lambda&\rightarrow \lambda\exp\{-i\A(t)\cdot(\R_i-\R_j)\}.
\end{align}
	On the diamond crystal structure, the $t$ hopping occurs between $A$ and $B$ sublattices, and the $\lambda$ term occurs between same-sublattice sites in neighboring unit cells.
	The lattice site positions can be further defined in terms of the unit cell position $\R_u$ and the sublattice position within the unit cell 
	${\mathbf{r}}_\tau,~\tau=A,B$, so that the Peierls term becomes
\begin{align}
	\R_i &\rightarrow \R_u + {\mathbf{r}}_\tau,\\
	\exp\{-i\A(t)\cdot(\R_i-\R_j)\} &\rightarrow\exp\{-i\A(t)\cdot(\R_u+{\mathbf{r}}_\tau-\R_{u'}-{\mathbf{r}}_{\tau'})\}.
\end{align}
	We can split Eq.~\ref{eq:hc} further into $\mathcal{H}_c=\mathcal{H}_{nn}+\mathcal{H}_{\mathcal{P},\mu}+\mathcal{H}_\text{soc}$, where, respectively, the first term is the nearest-neighbor hopping, the second term is the inversion symmetry-breaking term and chemical potential, and the third term is the spin-orbit coupling term. 
	The first and third terms are kinetic terms affected by the Peierls substitution.
	
	Applying the Fourier transform and Peierls substitution to $\mathcal{H}_{nn}$ and after some algebra, the nearest neighbor hopping term becomes
\begin{align}
		\mathcal{H}_{nn} &= \frac{t}{N_L}\sum_{\sigma{\mathbf{k}}}
	\left( e^{-i\A(t)\cdot{\mathbf{r}}_B} c_{A \sigma}^\dagger({\mathbf{k}}) c_{B \sigma}({\mathbf{k}}) \right.
		\{ 1 + e^{-i({\mathbf{k}}-\A(t))\cdot\a_1} + e^{-i({\mathbf{k}}-\A(t))\cdot\a_2}
		+ e^{-i({\mathbf{k}}-\A(t))\cdot\a_3}\}\\
		&\qquad\qquad+ e^{i\A(t)\cdot{\mathbf{r}}_B} c_{B \sigma}^\dagger({\mathbf{k}}) c_{A \sigma}({\mathbf{k}})
		\{ 1 +e^{i({\mathbf{k}}-\A(t))\cdot\a_1} + e^{i({\mathbf{k}}-\A(t))\cdot\a_2}
		+ e^{i({\mathbf{k}}-\A(t))\cdot\a_3}\} \left.\right),
\end{align}
where 
the $\a_i$ are the $fcc$ Bravais lattice vectors.
As a consequence of the two sublattice diamond crystal structure and their nonequivalence in the inversion symmetry broken case of $m\neq0$, the nontrivial phase factor cannot be absorbed away, and therefore the nearest-neighbor part of the Peierls transformation is not equivalent to a ${\mathbf{k}}\rightarrow {\mathbf{k}}-\A(t)$ transformation.
The nearest-neighbor term can be compactly written as
\begin{align}	
		\mathcal{H}_{nn}&=\frac{1}{N_L}\sum_{\sigma{\mathbf{k}}}
	\left( c_{A \sigma}^\dagger({\mathbf{k}}) c_{B \sigma}({\mathbf{k}}) u({\mathbf{k}},t) 
		+ c_{B \sigma}^\dagger({\mathbf{k}}) c_{A \sigma}({\mathbf{k}}) u^*({\mathbf{k}},t) \right),\\
		& u({\mathbf{k}},t)=t e^{-i\A(t)\cdot{\mathbf{r}}_B}\left(1 + \sum_{i=1}^3 e^{-i({\mathbf{k}}-\A(t))\cdot\a_i}\right).
\end{align}

Likewise, applying the Fourier transform and Peierls substitution to $\mathcal{H}_\text{soc}$, we get a conventional ${\mathbf{k}}\rightarrow {\mathbf{k}}-\A(t)$ type form:
\begin{align}
	\mathcal{H}_\text{soc} &= \sum_{\mathbf{k}} [ {\mathbf{D}}({\mathbf{k}},t)\cdot{\bm{\sigma}}_{\sigma\sigma'}c_{A\sigma}^\dagger({\mathbf{k}})c_{A'\sigma'}({\mathbf{k}}) - {\mathbf{D}}({\mathbf{k}},t)\cdot{\bm{\sigma}}_{\sigma\sigma'}c_{B\sigma}^\dagger({\mathbf{k}})c_{B'\sigma'}({\mathbf{k}}) ],
\end{align}
where the primed sublattice notation $A',~B'$ denotes the neighboring unit cell relative to sites $A,~B$ in the unit cell, ${\bm{\sigma}}_{\sigma\sigma'}$ is the vector of Pauli matrices ${\bm{\sigma}}=(\sigma_x, \sigma_y, \sigma_z)$, and the ${\mathbf{D}}({\mathbf{k}},t)$ vector components given by
\begin{align}
	D_x({\mathbf{k}},t) &= 4\lambda\sin(\tfrac{k_x-A_x(t)}{2})\left( -\cos(\tfrac{k_y-A_y(t)}{2}) + \cos(\tfrac{k_z-A_z(t)}{2}) \right),\\
	D_y({\mathbf{k}},t) &= 4\lambda\sin(\tfrac{k_y-A_y(t)}{2})\left( \cos(\tfrac{k_x-A_x(t)}{2}) - \cos(\tfrac{k_z-A_z(t)}{2}) \right),\\
	D_z({\mathbf{k}},t) &= 4\lambda\sin(\tfrac{k_z-A_z(t)}{2})\left( -\cos(\tfrac{k_x-A_x(t)}{2}) + \cos(\tfrac{k_y-A_y(t)}{2}) \right).
\end{align}

	Using the basis $\psi_{\mathbf{k}}=(c_{A \up},c_{B \up},c_{A \dn},c_{B \dn})^T$, $\psi_{\mathbf{k}}^\dagger$, the Bloch matrix $H_c({\mathbf{k}},t)$ for the conduction electron Hamiltonian $\mathcal{H}_c=\sum_{\mathbf{k}}\psi_{\mathbf{k}}^\dagger H_c({\mathbf{k}},t) \psi_{\mathbf{k}}$ is
\begin{align}
&H_c({\mathbf{k}},t) =\\
&\begin{pmatrix}
D_z({\mathbf{k}},t)+m-\mu & u({\mathbf{k}},t) & D_x({\mathbf{k}},t)-iD_y({\mathbf{k}},t) & 0 \\
u^*({\mathbf{k}},t) & -D_z({\mathbf{k}},t)-m-\mu & 0 & -D_x({\mathbf{k}},t)+iD_y({\mathbf{k}},t) \\
D_x({\mathbf{k}},t)+iD_y({\mathbf{k}},t) & 0 & -D_z({\mathbf{k}},t)+m-\mu & u({\mathbf{k}},t) & 0 \\
0 & -D_x({\mathbf{k}},t)-iD_y({\mathbf{k}},t) & u^*({\mathbf{k}},t) & D_z({\mathbf{k}},t)-m-\mu
\end{pmatrix}.
\end{align}

For completeness, we include the rest of the Bloch Hamiltonian, whose terms are not affected by the Peierls substitution.
The total Hamiltonian can be represented as $\mathcal{H}=\sum_{\mathbf{k}}\Psi_{\mathbf{k}}^\dagger H({\mathbf{k}},t) \Psi_{\mathbf{k}}$, where $\Psi_{\mathbf{k}}=(c_{A \up},c_{B \up},c_{A \dn},c_{B \dn},f_{A \up},f_{B \up},f_{A \dn},f_{B \dn})^T$ and similar for $\Psi_{\mathbf{k}}^\dagger$.
The hybridization sub-matrix is $rV\hat{I}$, where $\hat{I}$ is a $4\times4$ identity matrix, and $r$ is the auxiliary boson amplitude representing the hole fluctuations away from single particle per site $f$-electron occupation.
The $f$-electron sub-matrix is $(E_d+\ell-\mu)\hat{I}$, where $\ell$ renormalizes the localized $f$-level $E_d$.
Put together, the full $8\times8$ Bloch Hamiltonian matrix takes the form
\begin{align}
H({\mathbf{k}},t) =\begin{pmatrix}
	H_c({\mathbf{k}},t) & rV\hat{I} \\
	rV\hat{I} & (E_d+\ell-\mu)\hat{I}
\end{pmatrix}.
\end{align}
We have thus calculated the time-dependence of the system through the Peierls substitution of its Hamiltonian.

\subsection{Time evolution}\label{sec:time}

To capture the time evolution of the wavefunctions, we turn to the time-dependent Schr\"{o}dinger equation,
\begin{align}
	\frac{\partial}{\partial t} \Psi({\mathbf{k}},t) &= \frac{1}{i\hbar}H({\mathbf{k}},t)\Psi({\mathbf{k}},t),
\end{align}	
which is solved numerically by discrete time-evolution method. 
The wavefunction evolves by a unitary operator at each time step $\Delta t$,
\begin{align}
	 \psi({\mathbf{k}},t+\Delta t) &= \mathcal{U}(\Delta t)\psi({\mathbf{k}},t),\\
	 \mathcal{U}(\Delta t) &= e^{-i\mathcal{H}\Delta t}.
\end{align}
If the eigenenergies and eigenvectors of the system are computable, and the $j^{th}$ time-dependent wavefunction can be expanded as
\begin{align}
	 \psi_j({\mathbf{k}},t) &= \sum_n c_n(t) \phi_n({\mathbf{k}}),
\end{align}
with $\phi_n({\mathbf{k}})$ being the eigenfunctions with eigenenergies $\varepsilon_n({\mathbf{k}})$, then the time evolution can be represented as
\begin{align}
	 \psi_j({\mathbf{k}},t) &= e^{-i\mathcal{H} t} \psi_j({\mathbf{k}},0)\\
	&= \sum_n e^{-i\varepsilon_n({\mathbf{k}}) t} c_n(0) \psi_j({\mathbf{k}},0).
\end{align}

In equilibrium, at a time $t_{eq}=-\infty$ before a time-varying electric field pulse is applied, we obtain the eigenspectrum of the system as a set of eigenvectors $|\phi_n({\mathbf{k}},t_{eq})\ra$ and eigenenergies $\varepsilon_n({\mathbf{k}})$ where $n$ indexes the energy bands, obtained from the usual
\begin{align}
	\mathcal{H}_{ij}({\mathbf{k}},t_{eq})\phi_{j n}({\mathbf{k}},t_{eq}) &= \varepsilon_n({\mathbf{k}},t_{eq})\phi_{i n}({\mathbf{k}},t_{eq}).
\end{align}
where $i,j$ denotes the vector or matrix components.

The wavefunction at $t>t_{eq}$, using the quasiparticle basis weights, is obtained by the same unitary transformation,
\begin{align}
	|\phi_n({\mathbf{k}},t+\Delta t)\ra = \mathcal{U}(\Delta t) |\phi_n({\mathbf{k}},t)\ra,
\end{align}
where
\begin{align}
	\mathcal{U}_{ij}(\Delta t) &= e^{-i\mathcal{H}\Delta t}\\
	& = \phi_{in} e^{-i\varepsilon_n({\mathbf{k}})\Delta t} (\phi_{jn})^\dagger,
\end{align}
and the time increments $\Delta t$ are extremely small compared to the total time interval and pulse length.

\subsection{Current operator}\label{sec:current}

For a multi-orbital tight-binding model such as a heavy fermion system, or with different atoms occupying the unit cell, we can write a generic Hamiltonian
\begin{align}
	\mathcal{H} &= \sum_{ij,\alpha\beta} h_{ij,\alpha\beta} c_{i\alpha}^\dagger c_{j\beta} - \mu \sum_{i\alpha}c_{i\alpha}^\dagger c_{i\alpha}
\end{align}
where $c_{i\alpha}^\dagger,~ c_{i\alpha}$ are electron creation and annihilation operators at site $i$ ($j$), with a combined index $\alpha$ ($\beta$) for other degrees of freedom (spin, orbit, sublattice, etc.).
At this stage, the Hamiltonian describes both pristine and disordered systems.
The single particle hopping and on-site energy are described by $h_{ij,\alpha\beta}$, and the chemical potential by $\mu$.

We now define the electric polarization density operator
\begin{align}
	\hat{\vec{P}} &= \frac{e}{\mathcal{V}} \sum_i \R_i \hat{n}_i,
\end{align}
where $\R_i$ denotes the position of a lattice site, $\hat{n}_i=c_{i\alpha}^\dagger c_{i\alpha}$ is the electron number operator at site $i$, $V$ is the volume of the whole system, and $e$ is the electronic charge which we define to contain sign.
The charge current operator is defined as
\begin{align}
	\hat{\vec{J}} &= -\frac{i}{\hbar} [\hat{\vec{P}},\mathcal{H}]_-\\
	&= -\frac{ie}{\mathcal{V}\hbar} \sum_{ijl}\sum_{\alpha\beta\gamma} \R_i h_{jl,\beta\gamma}[c_{i\alpha}^\dagger c_{i\alpha}, c_{j\beta}^\dagger c_{l\gamma}],
\end{align}
where square brackets denote the commutator $[A,B]=AB-BA$.
The commutator can be worked out as:
\begin{align}
	[c_{i\alpha}^\dagger c_{i\alpha}, c_{j\beta}^\dagger c_{l\gamma}] &= c_{i\alpha}^\dagger[c_{i\alpha}, c_{j\beta}^\dagger c_{l\gamma}] + [c_{i\alpha}^\dagger, c_{j\beta}^\dagger c_{l\gamma}] c_{i\alpha}.
	%
\end{align}
Using the identity
\begin{align}
	[A, BC] = \{A, B\}C - B\{A, C\}
\end{align}
where the curly braces denote the anticommutator $\{A,B\}=AB+BA$, and using standard fermion anticommutator identities $\{c_\alpha^\dagger, c_\beta^\dagger \} =0= \{c_\alpha, c_\beta \}$ and $\{c_\alpha^\dagger, c_\beta \}=\delta_{\alpha\beta}$, this becomes
\begin{align}
	c_{i\alpha}^\dagger \{c_{i\alpha}, c_{j\beta}^\dagger \} c_{l\gamma} - c_{j\beta}^\dagger \{c_{i\alpha}^\dagger, c_{l\gamma} \}c_{i\alpha}
	&= c_{i\alpha}^\dagger c_{l\gamma} \delta_{ij}\delta_{\alpha\beta} - c_{j\beta}^\dagger c_{i\alpha} \delta_{il}\delta_{\alpha\gamma}.
\end{align}
The current operator then becomes
\begin{align}
	\hat{\vec{J}} &= -\frac{ie}{\mathcal{V}\hbar} \sum_{ijl}\sum_{\alpha\beta\gamma} \R_i h_{jl,\beta\gamma} \left( c_{i\alpha}^\dagger c_{l\gamma} \delta_{ij}\delta_{\alpha\beta} - c_{j\beta}^\dagger c_{i\alpha} \delta_{il}\delta_{\alpha\gamma} \right)\\
	&= -\frac{ie}{\mathcal{V}\hbar}\left[ \sum_{il}\sum_{\alpha\gamma} \R_i h_{il,\alpha\gamma} c_{i\alpha}^\dagger c_{l\gamma} - 
		\sum_{ij}\sum_{\alpha\beta} \R_i h_{ji,\beta\alpha} c_{j\beta}^\dagger c_{i\alpha} \right].
\end{align}
Exchanging the dummy indices in the first term $l\rightarrow j,~ \gamma\rightarrow\beta$, the operator is then
\begin{align}
	\hat{\vec{J}} &= -\frac{ie}{\mathcal{V}\hbar} \sum_{ij}\sum_{\alpha\beta} \R_i \left[ h_{ij,\alpha\beta} c_{i\alpha}^\dagger c_{j\beta}
	-h_{ji,\beta\alpha} c_{j\beta}^\dagger c_{i\alpha} \right],
\end{align}
and with a final exchange of dummy indices in the second term $j\leftrightarrow i,~ \beta\leftrightarrow\alpha$, we arrive at an expression for the current operator in terms of the displacement between pairs of lattice sites:
\begin{align}
	\hat{\vec{J}} &= -\frac{ie}{\mathcal{V}\hbar} \sum_{ij}\sum_{\alpha\beta} (\R_i-\R_j)  h_{ij,\alpha\beta} c_{i\alpha}^\dagger c_{j\beta}.
\end{align}

There are two gauge choices for the Fourier transform into momentum space:
\begin{align}	
	\text{I:}\qquad\qquad&c_{i\alpha}^\dagger(\R_i)= \frac{1}{\sqrt{N_L}}\sum_{\mathbf{k}} e^{-i{\mathbf{k}}\cdot\R_u}c_{{\mathbf{k}}\tau\alpha}^\dagger \\	
				&c_{i\alpha}(\R_i)= \frac{1}{\sqrt{N_L}}\sum_{\mathbf{k}} e^{ i{\mathbf{k}}\cdot\R_u}c_{{\mathbf{k}}\tau\alpha}\\
	\text{II:}\qquad\qquad&c_{i\alpha}^\dagger(\R_i)= \frac{1}{\sqrt{N_L}}\sum_{\mathbf{k}} e^{-i{\mathbf{k}}\cdot(\R_u+{\mathbf{r}}_\tau)}c_{{\mathbf{k}}\tau\alpha}^\dagger \\
				&c_{i\alpha}(\R_i)= \frac{1}{\sqrt{N_L}}\sum_{\mathbf{k}} e^{i{\mathbf{k}}\cdot(\R_u+\mathbf{r}_\tau)}c_{{\mathbf{k}}\tau\alpha},\\
\end{align}
where the total volume $\mathcal{V}=V_\Omega N_L$, the product of the unit cell volume with the number of unit cells.
We used gauge I to derive the Peierls substituted Hamiltonian in Sec.~\ref{sec:peierls}.
In general, the position of a lattice site $\R_i$ (site index) in a crystal structure can be expressed in terms of the location of the unit cell $\R_u$ and the relative position of the lattice site within the unit cell, also known as the lattice basis vector ${\mathbf{r}}_\tau$:
\begin{align}
	\R_i &= \R_u + {\mathbf{r}}_\tau.
\end{align}
In gauge II, the current operator can be Fourier transformed as
\begin{align}
	\hat{\vec{J}} &= -\frac{ie}{\mathcal{V}\hbar} \sum_{ij}\sum_{\alpha\beta} (\R_i-\R_j)  h_{ij,\alpha\beta} c_{i\alpha}^\dagger c_{j\beta}\\
	&= -\frac{ie}{V_\Omega\hbar} \sum_{uu'}\sum_{\tau\tau'}\sum_{\alpha\beta} (\R_u+{\mathbf{r}}_\tau - \R_{u'}-{\mathbf{r}}_{\tau'})h_{uu'\tau\tau',\alpha\beta}\\
		&\times\frac{1}{\sqrt{N_L}} \sum_{\mathbf{k}} e^{-i{\mathbf{k}}\cdot(\R_u+{\mathbf{r}}_\tau)} c_{{\mathbf{k}}\tau\alpha}^\dagger
		\frac{1}{\sqrt{N_L}} \sum_{{\mathbf{k}}'} e^{-i{\mathbf{k}}'\cdot(\R_{u'}+{\mathbf{r}}_{\tau'})} c_{{\mathbf{k}}'\tau'\beta}\\
	&= -\frac{ie}{V_\Omega N_L\hbar} \sum_{uu'}\sum_{\tau\tau'}\sum_{\alpha\beta} [(\R_u- \R_{u'})+({\mathbf{r}}_\tau -{\mathbf{r}}_{\tau'})]  
	 \sum_{{\mathbf{k}}{\mathbf{k}}'} e^{-i{\mathbf{k}}\cdot(\R_u+{\mathbf{r}}_\tau)} e^{-i{\mathbf{k}}'\cdot(\R_{u'}+{\mathbf{r}}_{\tau'})} c_{{\mathbf{k}}\tau\alpha}^\dagger c_{{\mathbf{k}}'\tau'\beta}.
\end{align}
We now redefine the coordinates to be relative to an origin unit cell position $\vec{R}_0$ (amounting to a change of index $u\rightarrow0, ~u'\rightarrow\nu$), so that the
the Hamiltonian matrix is relabeled as $h_{uu'\tau\tau',\alpha\beta}=h_{0\nu\tau\tau',\alpha\beta}$, and the displacement vectors become 
\begin{align}
	(\R_u- \R_{u'})+({\mathbf{r}}_\tau -{\mathbf{r}}_{\tau'}) &=(\R_0- \R_\nu)+({\mathbf{r}}_\tau -{\mathbf{r}}_{\tau'})\\
	&= -(\R_\nu+{\mathbf{r}}_{\tau'} -{\mathbf{r}}_\tau).
\end{align}
We multiply by $1=e^{i{\mathbf{k}}'\cdot\R_u}e^{-i{\mathbf{k}}'\cdot\R_u}$, and the sum over the exponentials becomes
\begin{align}
	&\sum_{{\mathbf{k}}{\mathbf{k}}'} e^{-i{\mathbf{k}}\cdot(\R_u+{\mathbf{r}}_\tau)}e^{i{\mathbf{k}}'\cdot(\R_{u'}+{\mathbf{r}}_{\tau'})}e^{i{\mathbf{k}}'\cdot\R_u}e^{-i{\mathbf{k}}'\cdot\R_u}\\
		&= \sum_{{\mathbf{k}}{\mathbf{k}}'} e^{-i{\mathbf{k}}\cdot(\R_u+{\mathbf{r}}_\tau)}e^{i{\mathbf{k}}'\cdot(\R_{u'}+{\mathbf{r}}_{\tau'}+\R_u-\R_u})\\
		&= \sum_{{\mathbf{k}}{\mathbf{k}}'} e^{-i({\mathbf{k}}-{\mathbf{k}}')\cdot\R_u} e^{i{\mathbf{k}}\cdot{\mathbf{r}}_\tau}e^{i{\mathbf{k}}'\cdot(\R_{u'}+{\mathbf{r}}_{\tau'}+\R_u)}\\
		&= N_L\sum_{{\mathbf{k}}{\mathbf{k}}'} \delta_{{\mathbf{k}}{\mathbf{k}}'} e^{i{\mathbf{k}}\cdot{\mathbf{r}}_\tau}e^{i{\mathbf{k}}'\cdot(\R_{u'}+{\mathbf{r}}_{\tau'}+\R_u)}\\
		&= N_L\sum_{\mathbf{k}} e^{i{\mathbf{k}}\cdot(\R_{u'}+{\mathbf{r}}_\tau-\R_u-\vec{r}_{\tau'})}\\
		&= N_L\sum_{\mathbf{k}} e^{i{\mathbf{k}}\cdot(\R_\nu+{\mathbf{r}}_\tau-{\mathbf{r}}_{\tau'})}.
\end{align}
With this, the current becomes
\begin{align}
	\hat{\vec{J}} &= \frac{ie}{V_\Omega\hbar} \sum_{{\mathbf{k}}\nu}\sum_{\tau\tau'}\sum_{\alpha\beta} (\R_\nu+{\mathbf{r}}_{\tau'} -{\mathbf{r}}_\tau) h_{0\nu\tau\tau',\alpha\beta} 
	 e^{i{\mathbf{k}}\cdot(\R_\nu+{\mathbf{r}}_\tau-{\mathbf{r}}_{\tau'})} c_{{\mathbf{k}}\tau\alpha}^\dagger c_{{\mathbf{k}}'\tau'\beta}.
\end{align}
The Hamiltonian matrix can be rewritten as
\begin{align}
	 \tilde{h}_{\tau\alpha,\tau'\beta}({\mathbf{k}}) = \sum_{\nu} h_{0\nu\tau\tau',\alpha\beta} e^{i{\mathbf{k}}\cdot(\R_\nu+{\mathbf{r}}_\tau-{\mathbf{r}}_{\tau'})} -\mu\delta_{\tau\tau'}\delta_{\alpha\beta},
\end{align}
so that the current operator is
\begin{align}
	\hat{\vec{J}} &= \frac{ie}{V_\Omega\hbar} \sum_{{\mathbf{k}}}\sum_{\tau\tau'}\sum_{\alpha\beta} (\R_\nu+{\mathbf{r}}_{\tau'} -{\mathbf{r}}_\tau) \tilde{h}_{\tau\alpha,\tau'\beta}({\mathbf{k}}) c_{{\mathbf{k}}\tau\alpha}^\dagger c_{{\mathbf{k}}'\tau'\beta}.\label{eq:current}
\end{align}
All quantities reported using this current expression are in units of $ie/V_\Omega \hbar$.

The ${\mathbf{k}}$-space gradient of the Hamiltonian matrix gives
\begin{align}
	\nabla_{\mathbf{k}} \tilde{h}_{\tau\alpha,\tau'\beta}({\mathbf{k}}) &= i\sum_\nu (\R_\nu+{\mathbf{r}}_{\tau'} -{\mathbf{r}}_\tau) h_{0\nu\tau\tau',\alpha\beta} e^{i{\mathbf{k}}\cdot(\R_\nu+{\mathbf{r}}_\tau-{\mathbf{r}}_{\tau'})},
\end{align}
	therefore,
\begin{align}	
	\hat{\vec{J}} &= \frac{e}{V_\Omega \hbar} \sum_{{\mathbf{k}}}\sum_{\tau\tau'}\sum_{\alpha\beta} [\nabla_{\mathbf{k}} \tilde{h}_{\tau\alpha,\tau'\beta}({\mathbf{k}})] c_{{\mathbf{k}}\tau\alpha}^\dagger c_{{\mathbf{k}}'\tau'\beta}.
\end{align}
In the quasiparticle basis
\begin{align}
		c_{{\mathbf{k}}\tau\alpha} = \sum_n \phi_{\tau\alpha n}({\mathbf{k}}) \gamma_{n{\mathbf{k}}}\label{eq:qp_1}\\
		c_{{\mathbf{k}}\tau\alpha}^\dagger = \sum_n \phi_{\tau\alpha n}^*({\mathbf{k}}) \gamma_{n{\mathbf{k}}}^\dagger\label{eq:qp_2},
\end{align}
the expectation value of the current is
\begin{align}	
	\langle\hat{\vec{J}}\rangle &= \frac{e}{V_\Omega\hbar} \sum_{{\mathbf{k}}}\sum_{\tau\tau'}\sum_{\alpha\beta}\sum_{nn'} [\nabla_{\mathbf{k}} \tilde{h}_{\tau\alpha,\tau'\beta}({\mathbf{k}})] \phi_{\tau\alpha n}^*({\mathbf{k}}) \phi_{\tau'\beta n'}({\mathbf{k}}) \langle\gamma_{n{\mathbf{k}}}^\dagger \gamma_{n'{\mathbf{k}}}\rangle\\
	&= \frac{e}{V_\Omega\hbar} \sum_{{\mathbf{k}}}\sum_{\tau\tau'}\sum_{\alpha\beta}\sum_{n} [\nabla_{\mathbf{k}} \tilde{h}_{\tau\alpha,\tau'\beta}({\mathbf{k}})] \phi_{\tau\alpha n}^*({\mathbf{k}}) \phi_{\tau'\beta n}({\mathbf{k}}) f(\eps_{n{\mathbf{k}}}),
\end{align}
where $f(\eps_{n{\mathbf{k}}})$ is the Fermi distribution function of the eigenenergies $\eps_{n{\mathbf{k}}}$.

We obtain the $i^{th}$ spatial component of the harmonic emission spectra in terms of the square of the Fourier transform of the dipole acceleration $\xi(\omega)=|\mathcal{FT}(d\langle J_i(t)\rangle /dt)|^2$ from Eq.~\ref{eq:current}.

\section{Results}\label{sec:results}

At this point we comment on the symmetry of the system and its impact on the low order harmonics of the current response.
This system resides on the diamond lattice and breaks inversion symmetry by differentiating the sublattices by the parameter $m$, equivalent to the zinc blende lattice structure of tetrahedral point group $T_d$, and space group \#216.
Since inversion symmetry is broken in both the WKSM and KI phases, generically even harmonics are expected to show up.
Interestingly, for this symmetry, transverse $1\omega_0$ responses proportional to $\sigma_{ij}$ for $i\neq j$ are expected to be zero, though later we find that the $1\omega_0$ response is present and rather robust.
Analyzing the second order susceptibility tensor from $P_i(2\omega)=\chi_{ijk}(2\omega;\omega,\omega)E_j(\omega)E_k(\omega)$ yields the same result as for the $0\omega$ response, that transverse components such as $\pm\chi_{xyz}=\chi_{yzx}=\pm\chi{zxy}$, with all other responses zero.
For the third order susceptibility tensor from $P_i(3\omega)=\chi_{ijkl}(3\omega;\omega,\omega,\omega)E_j(\omega)E_k(\omega)E_l(\omega)$, we find that for this point group the longitudinal responses ($i=j=k=l$) are nonzero, and the transverse components such as $\chi_{xyxy}, ~\chi_{yzyz}, ~\chi_{zxzx}$ are nonzero; however we find that generally we observe responses of the type $\chi_{zxyx}$ and $\chi_{zyxy}$. 

\subsection{Finite frequency high-harmonic response in linear polarization}\label{sec:omega_lin}

In what follows, we fix the following Hamiltonian parameters, corresponding to the valence regime of heavy fermions: $E_d=-4.45$, $V=3$, $\mu=-3.53022$, $r=0.69110$, $\ell=2.13821$, $k_B T=0.01$. 
The $r,~\ell,$ parameters are held fixed to their equilibrium values throughout the simulation.
We use a Monkhorst-Pack ${\bf{k}}$-mesh of $64\times64\times64$, and a time step of 0.001, corresponding to the femtosecond time scale.

\begin{figure}[h]
\centering
\includegraphics[width=12.0 cm]{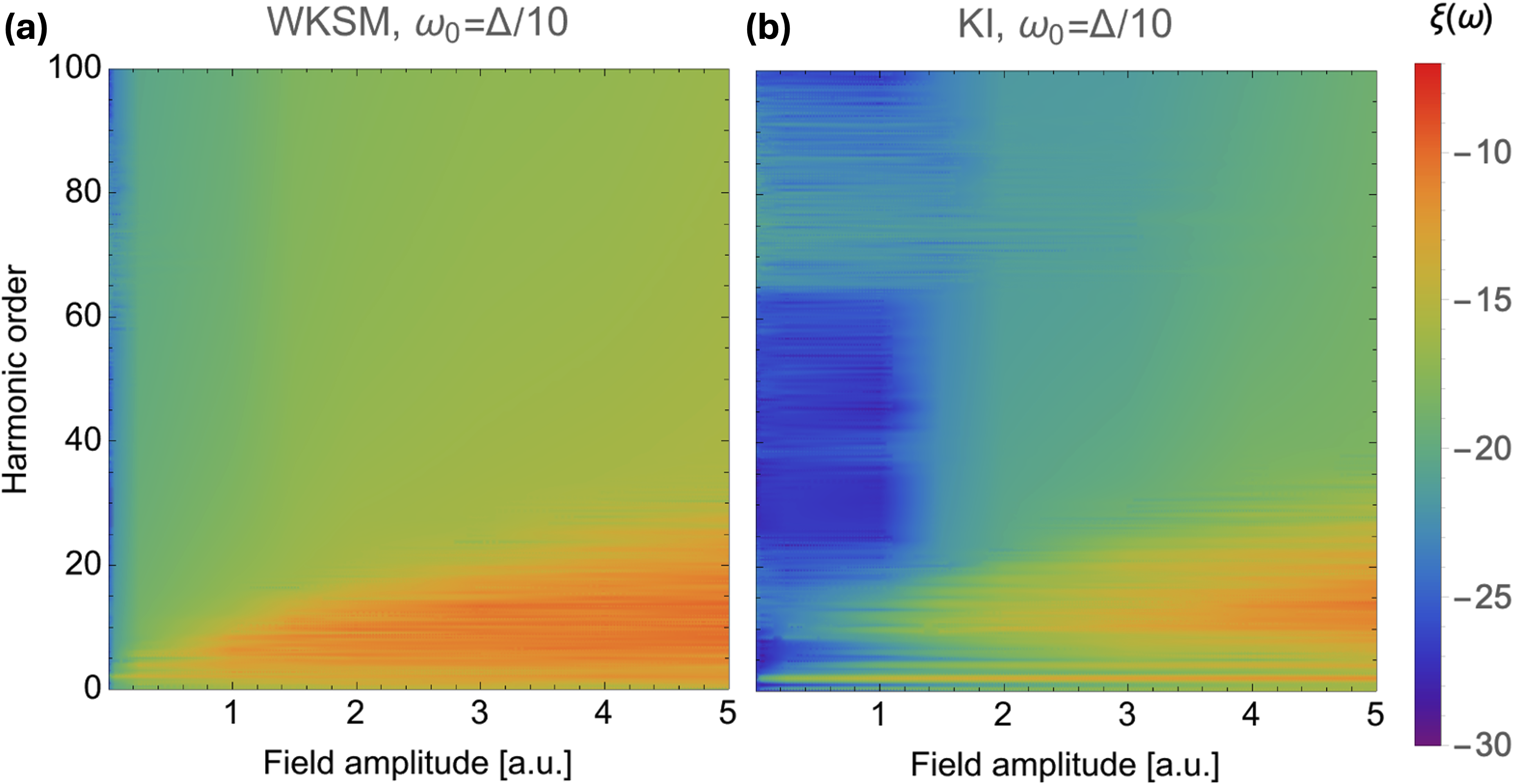}
\caption{Contour plots of harmonic intensity spectra $\xi(\omega)=\log|\mathcal{FT}(d\langle J_z(t)\rangle/dt)|^2$ in terms of harmonic orders $n\omega_0$ versus field amplitude $|A_0|$, using an input frequency of $\omega_0=\Delta/10$ compared to the Kondo insulator gap $\Delta$, corresponding to low terahertz frequencies.
(\textbf{a}) Weyl-Kondo semimetal harmonic spectra.
(\textbf{b}) Kondo insulator harmonic spectra.
\label{fig:colormap}}
\end{figure} 

We first use a finite frequency ultrafast pulsed source to excite the WKSM and KI systems to contrast their HHG spectra, where the pulse envelope for each component of ${\bf{A}}(t)$ is:
\begin{align}
A_i(t)=\frac{A_0}{\omega_0}\sin^2\left(\frac{\omega_0 t}{2N_0}\right)\sin(\omega_0 t)\Theta(-t)\Theta\left(t+\frac{2\pi N_0}{\omega_0}\right),
\end{align}
where $N_0=5$ is the number of cycles used, the pulse width is $2\pi N_0/\omega_0$, $\omega_0$ is the driving frequency, and $\Theta$ is the Heaviside-Theta function.
Linearly polarized light along $\frac{1}{\sqrt{2}}(\hat{x}+\hat{y})$ and frequency $\omega_0=\Delta/10$ in terms of the KI gap $\Delta$ is used, and we calculate the transverse current component $J_z(t)$.
Since the electric dipole acceleration is proportional to $\xi(\omega)=\log|\mathcal{FT}(d\langle J_z(t)\rangle/dt)|^2$, we calculate and analyze this quantity as the HHG intensity spectra.
We use a 5-cycle pulse width, and simulate in the time domain until a steady state is reached.

Fig.~\ref{fig:colormap} shows contour plots of the harmonic intensity spectra $\xi(\omega)$ for varying driving field intensities and harmonic orders, for a driving field corresponding to low terahertz frequencies. 
The WKSM displays a long-lived and noisy spectrum compared to the KI, with a continuum of low order harmonics compared to the low intensity of KI harmonic orders in the 4-10$\omega_0$ range, corresponding to the Kondo gap between the bands around the Fermi energy.
The higher intensity response of low order harmonics can be attributed to the readily available interband transitions existing for the Weyl cones of the WKSM.
Thus the analysis in Fig.~\ref{fig:colormap} indirectly images the linear dispersion of the WKSM and the KI gap.

\begin{figure}[h]
\centering
\includegraphics[width=\textwidth]{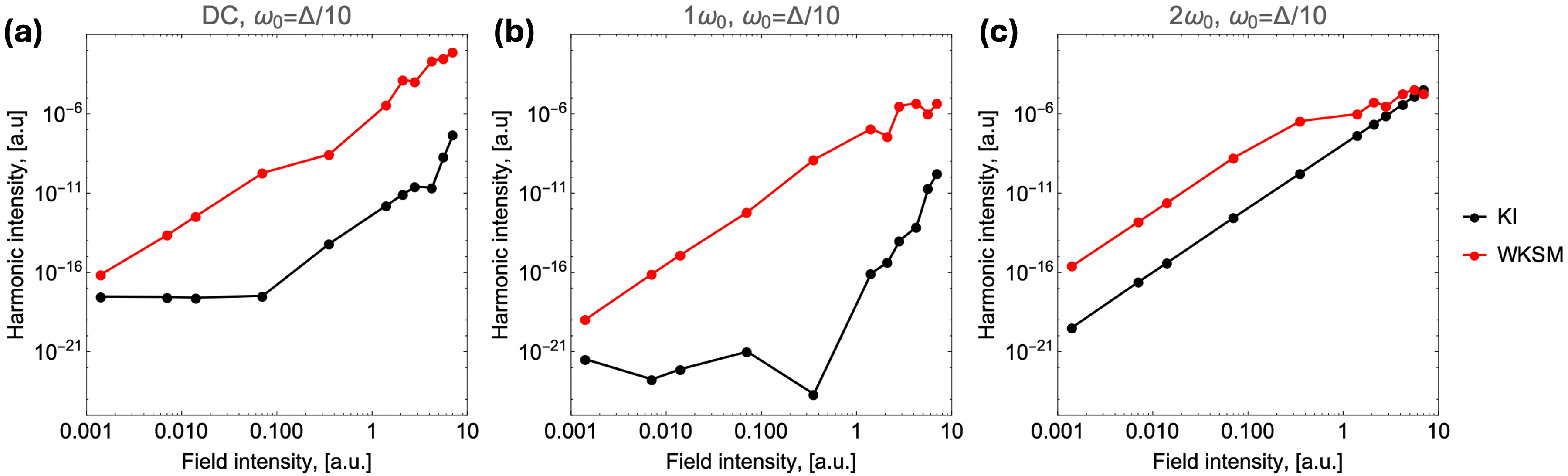}
\caption{Peaks of harmonic spectral intensity $|\mathcal{FT}(d\langle J_z(t)\rangle/dt)|^2$ versus field amplitude $|A_0|$, comparing the Weyl-Kondo semimetal (red) to the Kondo insulator (black), where the input frequency is $\omega_0=\Delta/10$ compared to the Kondo insulator gap $\Delta$.
(\textbf{a}) DC ($\omega=0\omega_0$ current rectification) peak responses. 
(\textbf{b}) $\omega=1\omega_0$ peak responses. 
(\textbf{c}) $\omega=2\omega_0$ peak responses.
\label{fig:peaks}}
\end{figure} 

Fig.~\ref{fig:peaks} shows the peaks of harmonic intensity of the data of Fig.~\ref{fig:colormap} for the 0$\omega_0$, 1$\omega_0$, and 2$\omega_0$ peaks over a range of driving field amplitudes $|A_0|$, for the same driving frequency $\omega_0=\Delta/10$.
In red, the WKSM displays a consistently higher harmonic intensity response, save for the higher field intensity tail of the $2\omega_0$ response.
In terms of the DC response, the WKSM is far more efficient in converting the finite $\omega_0$ field excitation to a $0\omega_0$ response compared to the KI. 
The WKSM displays higher field intensity deviations from power-law behavior, entering into the nonperturbative regime, whereas the KI (in black) shows only power law dependence in its 2$\omega_0$ response.
In the DC and 1$\omega_0$ spectral intensity, the KI shows a threshold behavior that is nearly flat at low field intensities and rises irregularly at high field intensities.

We note that the 0$\omega_0$ response is only present in the $z$-component of the current, and not in the $x$ or $y$-component. 
Since it shows up more strongly in the WKSM, especially at low field intensity, we can attribute this to the Weyl nodes of the WKSM.
The 0$\omega_0$ response present in the KI is a result of photoinduced metallicity, where a certain strength of the driving field is required to excite electrons across the gap $\Delta$. 
We emphasize that the $J_z$ component responses are driven by noncentrosymmetry.

\subsection{Finite frequency high-harmonic response in circular polarization}\label{sec:omega_circ}

The parameters used in this subsection are $E_d=-7.0$, $V=9.28571$, $t=1$, $\lambda=0.5$,  $k_B T=0.01$, and $\mu$, $r$, $\ell$ are solved self-consistently at each $m$. 
The $r,~\ell,$ parameters are held fixed to their equilibrium values throughout the simulation.
The $\mathbf{k}$-mesh is $48\times48\times48$, time step 0.001.

\begin{figure}[h]
\centering 
\includegraphics[width=\textwidth]{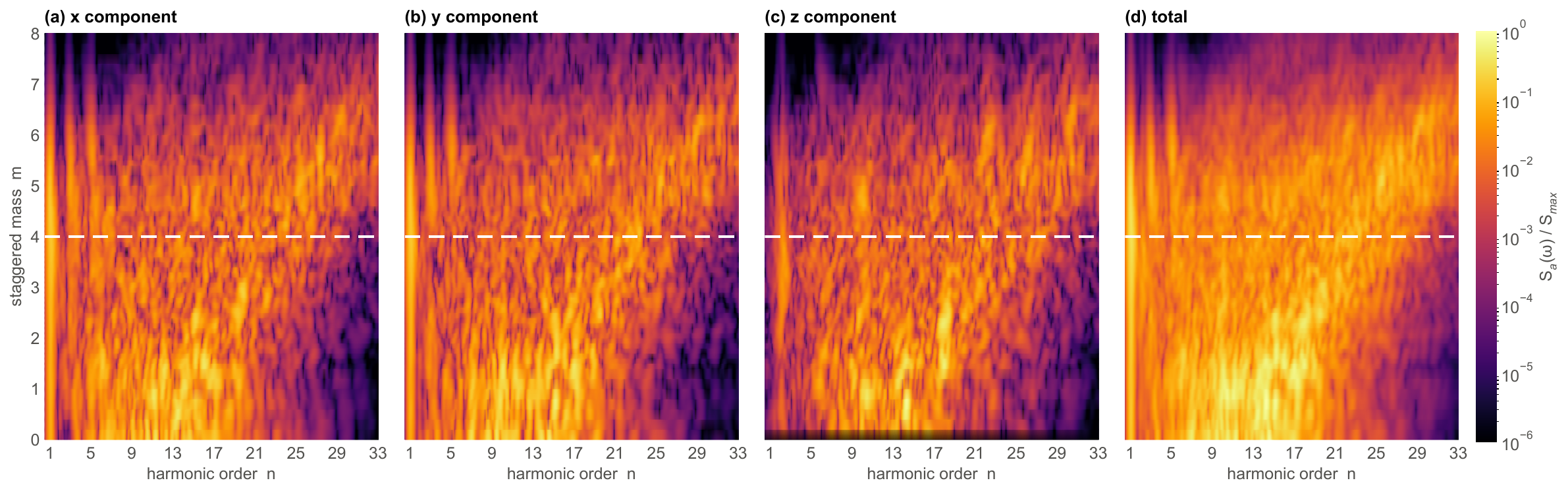}
\caption{High-harmonic spectra $\xi(\omega)$ as a function of topological tuning parameter $m$ (vertical axis) and harmonic order (horizontal axis) using xy-plane circularly polarized light of frequency $\omega_0=\Delta/4$.
(\textbf{a})  x-component, (\textbf{b})  y-component, (\textbf{c})  z-component, (\textbf{d})  total intensity.
 \label{fig:xyzt}}
\end{figure}   
\unskip

In Fig.~\ref{fig:xyzt}, we excite the system with right-handed circularly polarized light in the xy-plane with a frequency of $\omega_0=\Delta/4$ compared to the KI gap $\Delta$, when $m=2$, $\lambda=0.01$, at a maximum light intensity of $A_0=10$ (arbitrary units) and  circularly polarized pulse shape:
\begin{align}
    \mathbf{A}(t) = A_0 \sin^2\left(\frac{\omega_0 t}{2N_0}\right) (\sin \omega_0 t, \cos \omega_0 t, 0).
\end{align}
On the vertical axis, we sweep the $m$ parameter, keeping $\lambda=0.5$ fixed, taking the system's state from the WKSM phase (below the dashed line) to the KI phase (above the dashed line marking $m/8\lambda=1$).
The range of harmonic responses is displayed on the horizontal axis out to the 33$^\text{rd}$ harmonic.

Consistent in all four plots is the absence of strong intensity in higher harmonics in the WKSM phase, at the lower right region of Fig.~\ref{fig:xyzt}(a)-(d).
This region shrinks as the topological phase transition (dashed line) is approached and disappears nearly at the phase transition point.
In contrast, in the WKSM phase, a region of high intensity begins in the 7$^\text{th}$-17$^\text{th}$ harmonic range near $m\sim0$ and progresses to higher harmonics and a narrower harmonic range as the topological transition is approached (this is most apparent in Fig.~\ref{fig:xyzt}(d)).
The brightest harmonics persist in sliding into a higher harmonic range across the transition through the KI phase but lessen in intensity after the critical point, ending in the upper 22$^\text{nd}$-30$^\text{th}$ harmonic range.

\begin{figure}[h]
\centering
\includegraphics[width=\textwidth]{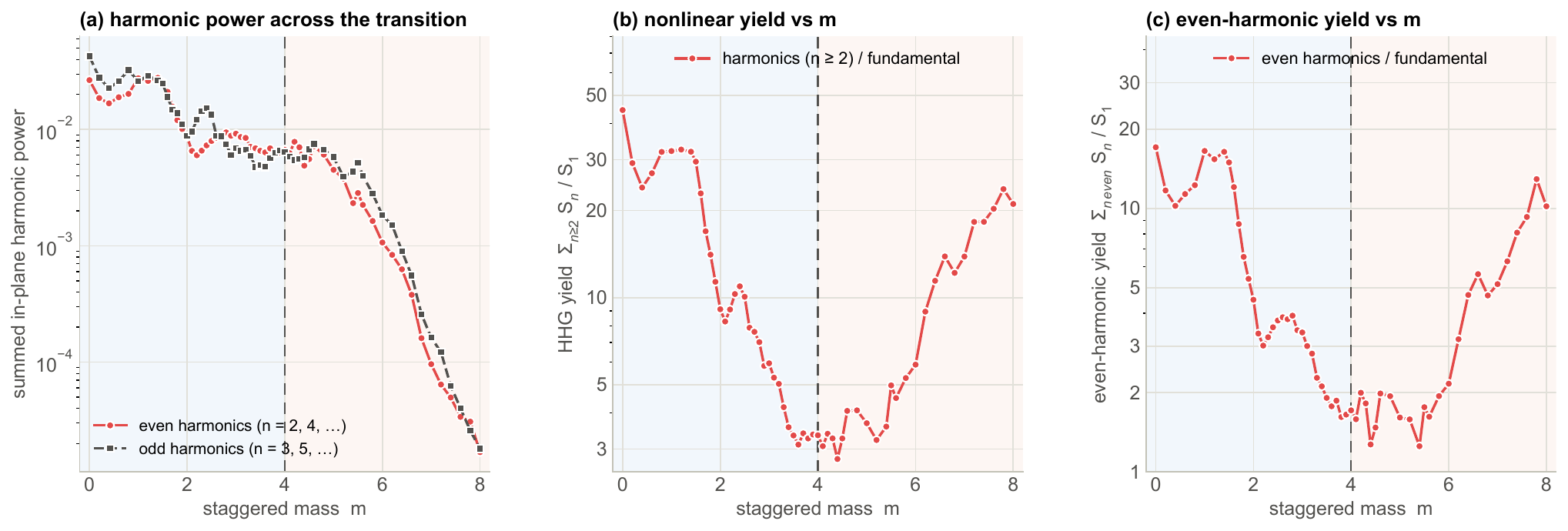}
\caption{ Harmonic power and efficiency across the topological phase transition (vertical dashed line), tuned by topological tuning parameter $m$, which is physically related to the amount of on-site potential difference, and breaks inversion symmetry in the diamond crystal structure.
(\textbf{a}) Harmonic power of the even harmonics in red dots and odd harmonics in gray squares. Lines are a guide to the eye.
(\textbf{b}) Nonlinear efficiency as a function of $m$, calculated as the harmonic power above the 1$^\text{st}$ (linear) harmonic.
(\textbf{c}) Even-harmonic yield versus $m$ normalized to $s(\omega_0)$. 
\label{fig:harmeff}}
\end{figure} 

Fig.~\ref{fig:harmeff}(a)-(c) show the harmonic power and efficiency as a function of the topological tuning parameter $m$, with the transition at $m=4$ marked by a vertical dashed line.
In the blue shaded region to the left of each figure is the WKSM phase, and on the right (shaded orange) is the KI phase.
Fig.~\ref{fig:harmeff}(a) demonstrates that the harmonic power shows a gradual drop as the Weyl nodes emerge from the Dirac point at $X$ when $m=0$, across the Brillouin zone boundary to the $W$ point where nodes of opposite chirality annihilate at $W$ when $m=4$ and open a topologically trivial gap in the KI phase.
At around $m=4.3$ ($\Delta=\omega_0$), the even harmonics begin to decrease more steeply, while the odd harmonics begin to decrease steeply at about $m=5.4$ ($\Delta=4.6\omega_0$).
This indicates that at least the even harmonic power is related to the gap size.

In Fig.~\ref{fig:harmeff}(b), the HHG yield versus staggered mass shows a large decrease within the WKSM phase, then is relatively flat, and begins to increase in the KI phase.
The nonlinear conversion efficiency is the light generated at new frequencies divided by the light returned at the driving frequency:
\begin{align}
\eta_\text{eff}&=\frac{\sum_{n\geq 2}s(n\omega_0)}{s(\omega_0)}.
\end{align}
$s(n\omega_0)$ is the integrated in-plane emission weight around the $n$-th harmonic, calculated by Fourier-transforming the Hann-windowed current during the pulse and summing the radiated power over a $\pm0.4\omega_0$ band, rather than measuring a single peak height:
\begin{align}
s(n\omega_0)&=\int_{(n-0.4)\omega_0}^{(n+0.4)\omega_0}\omega^2\left(|J_x(\omega)|^2+|J_y(\omega)|^2\right)d\omega .
\end{align}
In Fig.~\ref{fig:harmeff}(c) only the even harmonics are selected from the nonlinear conversion efficiency, with a similar behavior displayed.
The harmonic response appears to track whether the driving frequency can detect that the gap has opened rather than the proximity of the Weyl nodes, as it is a continuous signature rather than a switch at the critical mass.

\subsection{DC response}\label{sec:omega_DC}

In this subsection we explore the instantaneous DC ($\omega_0=0$) excitation and current response $\log|J_y|$ using Eq.~\ref{eq:current}.
We excite with a uniform DC electric field oriented in various high-symmetry directions of the underlying crystal structure, and contrast the current responses of the WKSM and KI phases.
The following parameters were used for this subsection: $E_d=-4.45$, $V=3$, $\mu=-3.53022$, $r=0.69110$, $\ell=2.13821$, $k_B T=0.01$.

\begin{figure}[h]
\centering 
\includegraphics[width=\textwidth]{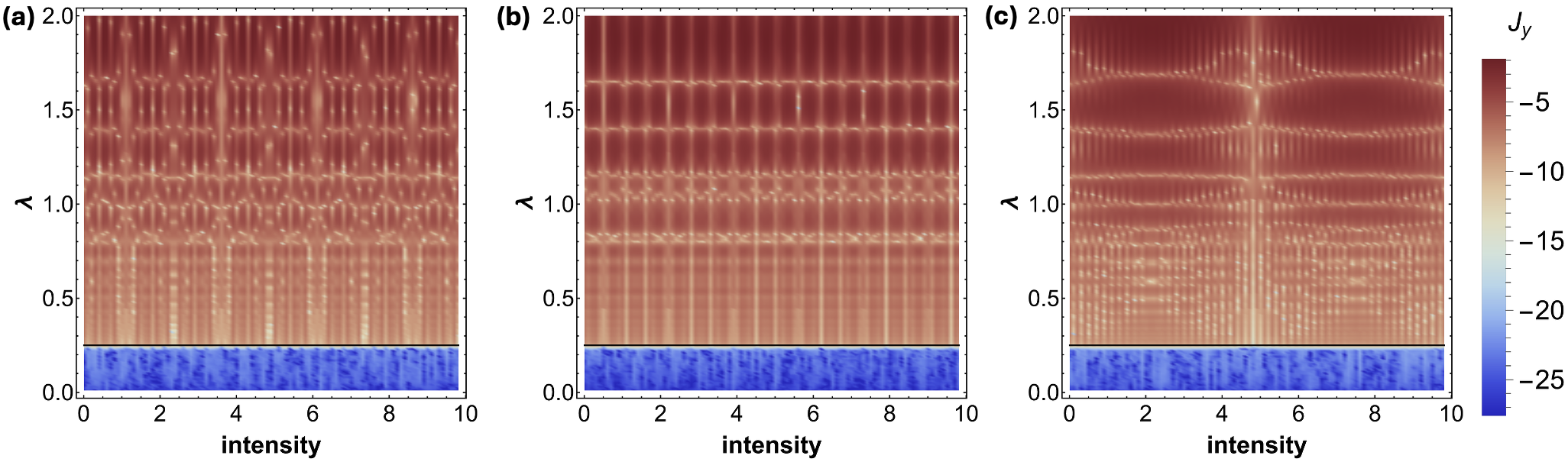}
\caption{DC current $\log J_y$ as a result of DC field excitation in varying directions versus field intensity and spin-orbit coupling $\lambda$ indicated on the vertical axis. 
The horizontal line marks the topological phase transition between WKSM (above) and KI (below). Here the staggered mass $m=2$.
(\textbf{a}) $K$-direction excitation along $(3\pi/2,3\pi/2,0)$.
(\textbf{b}) $L$-direction excitation along $(\pi,\pi,\pi)$.
(\textbf{c}) $X$-direction excitation along $(0,2\pi,0)$.
\label{fig:dc3}}
\end{figure}

As shown in Fig.~\ref{fig:dc3}, there is a striking difference in the DC response of the KI versus WKSM.
We show that for three different directions of field excitation in the Brillouin zone ($K,L,X$) that the WKSM response tends to exceed $J_y\propto 10^{-15}$, while the KI response is strongly suppressed at $J_y\propto 10^{-20}$, accompanied by a sharp jump at the phase transition when the Weyl nodes annihilate and produce the Kondo gap.
The DC field, for a range of excitation field intensities (as shown in the horizontal axes), is able to distinguish between the gapped and gapless phases across the topological phase transition.

\begin{figure}[h]
\centering 
\includegraphics[width=\textwidth]{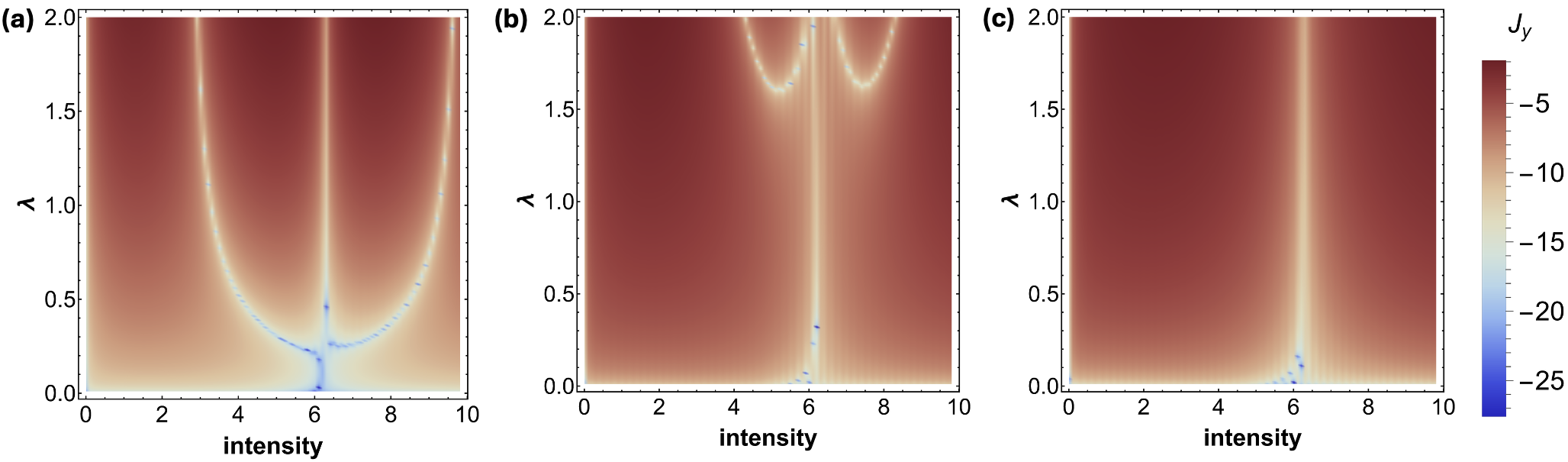}
\caption{DC current $\log J_y$ as a result of DC field excitation in the $X$ direction versus field intensity and inversion symmetry breaking parameter $m$ at a fixed topological tuning ratio $m/8\lambda$ with the spin-orbit coupling $\lambda$ indicated on the vertical axis. Topological tuning ratio is fixed at:
(\textbf{a}) $m/8\lambda=0.125$,
(\textbf{b}) $m/8\lambda=0.5$,
(\textbf{c}) $m/8\lambda=2$.
\label{fig:dc}}
\end{figure}   

We show in Fig.~\ref{fig:dc} the DC field-driven current response of the WKSM (\textbf{a})-(\textbf{b}) compared to that of the KI (\textbf{c}) for fixed ratios of $m/8\lambda$, with the field polarized in the $y$-direction, and plot the $\log$ of the current as a function of excitation field intensity and inversion symmetry breaking parameter $m$, while keeping the ratio of $m/8\lambda$ fixed so that the phase is consistent throughout the plot.
The $m/8\lambda=0.125$ and $m/8\lambda=0.5$ ratio fixes the WKSM phase, while the $m/8\lambda=2$ fixes the KI phase.
In the WKSM response, we can see that a $\Psi$-shaped feature of current dip is evident, while in the KI phase, the shape has transitioned to a vertical line of small current response occurring at a field intensity of about 6 (arbitrary units).

\subsection{Time-dependence of auxiliary boson parameters}\label{sec:rt_lt}

In this section we allow the self-consistently determined parameters $r, ~\ell$ to evolve as a function of time in response to a pulse of $\omega_0$ field polarized in the $x$-direction for the WKSM ($m=1, ~\lambda=0.5$) in the Kondo regime of $E_d=-6.5$, $V=7.05$, $\mu=-9.39337$, $r(t=0)=0.25978$, $\ell(t=0)=6.85708$, $k_B T=0.01$.
The pulse
\begin{align}
    A_x(t) &= 2\pi A_0 \exp\left(-\frac{(t-t_c)^2}{2 t_b^2}\right)\cos(\omega_0(t-t_c))
\end{align}
is centered at $t_c=-5$ (in units roughly corresponding to the femtosecond scale) and terminates before $t=0$, with the maximum pulse strength $A_0=0.159$ and a full-width half-maximum of $t_b=1$.
The time step used is 0.001, with a ${\bf{k}}$-mesh of $24\times24\times24$.
We approximate $\mu$ as fixed as a function of time, as the time scales are more on the order of the electronic relaxation time than the thermodynamic time scale.

Intuitively, the auxiliary boson parameter $r$ renormalizes the hybridization term and thus the bands around the Fermi energy, which is where the Weyl nodes in the WKSM phase are Kondo-pinned.
The Fermi velocity around the Weyl nodes is therefore strongly renormalized in WKSMs to be ultraslow.
The quantity $r^2$ can be interpreted as the hole fluctuation away from half-filling of the $f$-electron orbitals; double occupation is projected out in the $U\rightarrow\infty$ limit.
The $\ell$ parameter is a Lagrange multiplier enforcing the total filling of the $f$-electrons and acts as a local chemical potential-like correction to the $f$-electrons' localized level.

\begin{figure}[h]
\centering
\includegraphics[width=\textwidth]{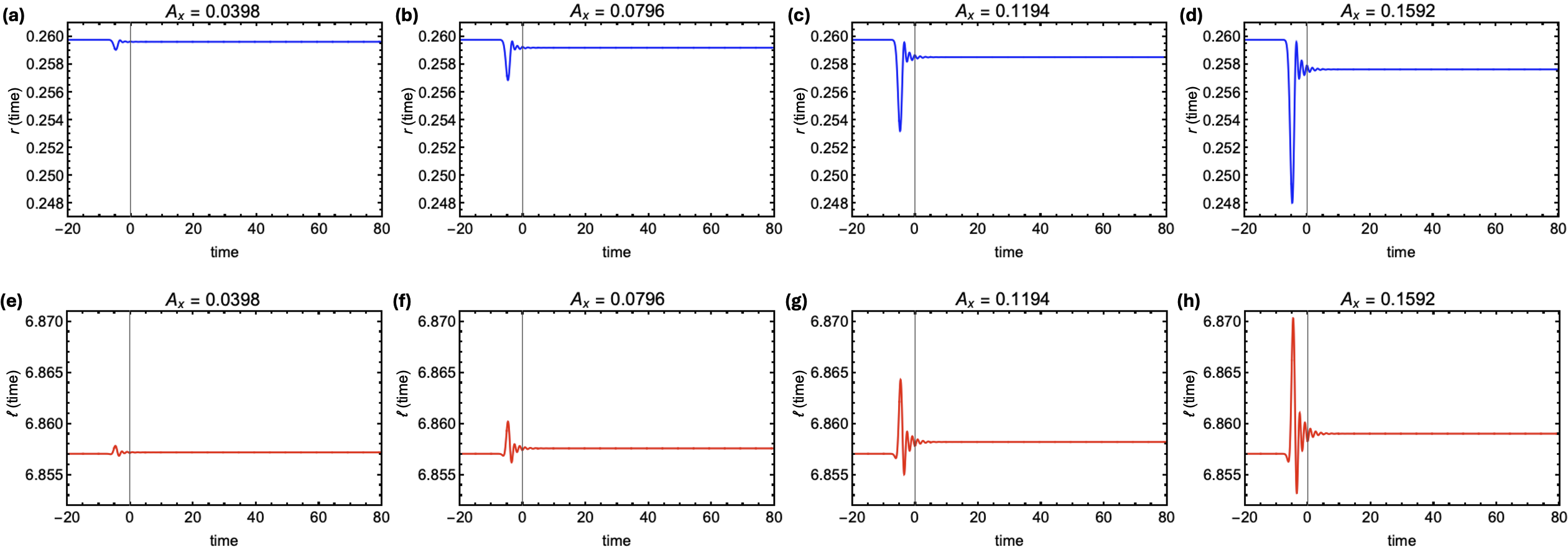}
\caption{Evolution of auxiliary boson parameters as a function of time in the Weyl-Kondo semimetal at four field amplitudes. An $\omega_0=0$ excitation pulse in the $x$-direction is used. {\bf{(a)}}-{\bf{(d)}}: $r(t)$ as in Eq.~\ref{eq:roft}. {\bf{(e)}}-{\bf{(h)}}: $\ell(t)$ as in Eq.~\ref{eq:loft}. \label{fig:flt}}
\end{figure}

The Eqs.~\ref{eq:sc_r}-\ref{eq:sc_l} can be inverted and put in terms of Peierls-substituted time-dependent quantities to yield the following time-dependent expressions for $r(t)$ and $\ell(t)$:
\begin{align}
	 r(t) &= \sqrt{1-\frac{1}{2N_u}\sum_{n,{\mathbf{k}}} \sum_{i=5}^{i=8} \langle\phi_{in}^*({\mathbf{k}},t) f[\eps_{n{\mathbf{k}}}]\phi_{in}({\mathbf{k}},t)\rangle},\label{eq:roft}\\
	\ell(t) &= -\frac{V}{2rN_u}\sum_{n,{\mathbf{k}}}\sum_{i=1}^{i=4} f[\eps_{n{\mathbf{k}}}]\langle \phi_{in}^*({\mathbf{k}},t) \phi_{(i+4)n}({\mathbf{k}},t)\rangle\label{eq:loft}
\end{align}
These quantities are calculated in Fig.~\ref{fig:flt}.

\unskip

We track the time evolution of the auxiliary boson order parameters with the application of an $\omega_0=0$ pulse polarized in the $x$-direction, along $X$ of the Brillouin zone, with the pulse ending at time=0.
As can be seen in Fig.~\ref{fig:flt}{\bf{(a)}}-{\bf{(h)}}, the pulse causes the auxiliary boson parameters to fluctuate in response from their initial equilibrium values, then settle at steady-state values that are suppressed (in the case of $r(t)$) or enhanced (in the case of $\ell(t)$) from the initial values.
The pulse has effectively caused the state of the system to undergo a change into another post-pulse state.

\section{Discussion}\label{sec:dis}

First, HHG largely probes the difference in the band structure of the two states.
In the linear polarization study of Fig.~\ref{fig:colormap}, the HHG spectrum of the WKSM is consistent with a linearly dispersing semimetallic band structure, while that of the KI is consistent with a system that has a small gap at the Fermi energy.
In terms of excitation frequency, the excitation energy is well below that of the gap ($\omega_0=\Delta/10$), so the HHG is able to resolve small interband transitions between bands that are close together.
The analysis of this data in Fig.~\ref{fig:peaks} shows confirmation that the overall intensity of the HHG spectrum is weaker in the KI than the WKSM, except the 2$^\text{nd}$ harmonic at very high field intensity.
This can be attributed to interband coherences which yield a photo-induced metallicity in the KI.
In the circular polarization study of Fig.~\ref{fig:xyzt}, we found clear differences in the intensity pattern between the WKSM and KI, which nonetheless transition smoothly across the topological phase transition.
Across this transition, the harmonic power begins to sharply decrease for the KI state when the gap opens past the excitation frequency energy threshold, but not as Weyl nodes annihilate at the critical staggered mass.
This indicates that the change in HHG yield are model-dependent band structure responses.

Second, in the DC response, we see a marked difference in the current produced in the WKSM state versus the KI state.
The DC response is practically nonexistent in the KI when spin-orbit coupling is reduced beyond the phase transition threshold.
The spin orbit coupling versus incident DC field strength also displays qualitative differences between the WKSM and KI phases, as a low-current $\Psi$-shaped feature is present in the WKSM and not in the KI.
This provides a useful check on the previous results, that the DC response can distinguish between the semimetal and gapped states.
Contrasting with the finite-$\omega_0$ results, the KI needs a finite frequency and large intensity of incident field to appreciably produce a finite 0$\omega_0$ response.
The DC response also increases with increasing spin-orbit coupling strength.

Third, in the pulsed DC response of the time-dependent auxiliary boson parameters, we observed an induced small change in the post-pulse state of the WKSM.
This shows that although the heavy-fermion state is incredibly robust in the WKSM, it is possible to change the quantitative features of that state.
The quantity $r^2$ characterizes the hole-fluctuation away from single occupation of the $f$-electron on a lattice site, and is roughly proportional to the conduction electron filling of the system $n_c$, whereas the $\ell$ Lagrange multiplier characterizes the enforcement of the $f$-electron filling, roughly proportional to $n_f=1-r^2$.
The total filling $n_\text{tot}=n_c+n_f=1$ per site stays fixed throughout the time-dependent simulation.
The pulse thus influences the relative densities of $f$-electrons and conduction electrons, influencing the heavy fermion order parameters, which could feasibly tune the system away from the mixed valence regime.
This gives a strong indication that photo-induced phase transitions may possibly drive the system to Kondo destruction where $r(t=\infty)\rightarrow0$ if a stronger or finite frequency pulse is used.
The result may indicate that control of post-pulse states using pulsed THz sources are possible.

Since the leading WKSM candidate material is Ce$_{3}$Bi$_{4}$Pd$_{3}$, and is closely related to the well-known KI Ce$_{3}$Bi$_{4}$Pt$_{3}$, one can ask whether THz HHG spectroscopy is possible across the topological phase transition.
The series Ce$_{3}$Bi$_{4}$(Pt$_{1-x}$Pd$_x$)$_{3}$ has been explored before, finding that the degree of spin orbit coupling is tuned without changing the lattice volume~\cite{Dzsaber2017}.
The specific heat data there also indicate that the Weyl linear dispersion evolves under $T\leq T_K=13$K, which poses a challenge to perform HHG under cryostat conditions, but there have been HHG experiments done on the high-temperature superconductor YBCO down to about 88K; future experimental setups may be able to push the temperature down to Kondo temperatures~\cite{Alcala2022}.
Our model uses a mean-field Anderson lattice model on the diamond crystal structure, which has a different space group than that of Ce$_{3}$Bi$_{4}$Pt$_{3}$, so although qualitative comparisons can be made, no quantitative comparisons are attempted.

In conclusion, we calculated the HHG spectroscopy, DC response, and photo-induced post-pulse states in a well-controlled model of a topological phase transition between WKSM and KI states.
In this work, we have developed the formalism to accommodate a finite field on the WKSM and KI model.
We have found that for the diamond crystal structure, an extra phase term affects the kinetic hopping term, which modulates the bandwidth about the Fermi energy as the field changes dynamically.
This is likely to affect other crystal lattice systems with multiple sites per unit cell in the same treatment.
We also advanced an expression for the non-perturbative current based on the polarization equation of motion, which contains all orders of the electric field and only depends on the lattice basis, multi-orbital Hamiltonian, and quasiparticle basis functions (Eq.~\ref{eq:current} with Eqs.~\ref{eq:qp_1}-\ref{eq:qp_2}).
This covers all orders of HHG spectroscopy in a complex model system.
Other approaches, such as separating out the inter- and intraband contributions or expanding perturbatively in orders of the excitation field are only practical on simple models or low harmonic orders.

Though there are claims that no universal signatures of topology exist~\cite{Neufeld2023}, and that for Weyl semimetal systems with pairs of nodes related by symmetry and at the Fermi energy (as in the WKSM), HHG is largely reflective of crystal symmetry group-dictated selection rules~\cite{Zhang2024}, the band structure at small energy resolutions can yet be distinguished between the WKSM and KI states.

%
%

\textbf{Acknowledgments:} {The authors gratefully acknowledge Silke Paschen and Chris Lane for illuminating conversations.}

\textbf{Funding:} {This work was performed in part at Aspen Center for Physics, which is supported by National Science Foundation grant PHY-2210452.
This work was funded through the Cal. State. Long Beach and the Ohio State University Partnership for Education and Research in Topological Materials, a National Science Foundation PREM, under Grant No. 2425133.
This work was supported by the Klaus-Tschira Stiftung under Grant No. 00.025.2025.
The work at Los Alamos was carried out under the auspices of the U.S. Department of Energy (DOE) National Nuclear Security Administration (NNSA) under Contract No. 89233218CNA000001, and in part by Center for Integrated Nanotechnologies, a DOE BES user facility, in partnership with the LANL Institutional Computing Program for computational resources.}






\printbibliography

\end{document}